\documentclass[1p,review]{elsarticle}
\pdfoutput=1
             
\usepackage{url}

\usepackage[hidelinks,colorlinks=true,breaklinks]{hyperref}
\usepackage{breakurl}

\usepackage{subcaption}
\usepackage{wrapfig}
\usepackage{xcolor}
\usepackage{lineno}
\usepackage{natbib}
\usepackage{amsmath}
\usepackage{siunitx}

\let\origautoref\autoref
\def\autoref#1{{\origautoref{#1}}}
\DeclareSIUnit\photoelectron{pe}

\usepackage[final]{changes}  

\title{Validation and integration tests of the JUNO
20-inch PMTs readout electronics}
\author[a]{Vanessa~Cerrone}
\author[a,b]{Katharina~von~Sturm\corref{cor1}}
\cortext[cor1]{Corresponding author}
\author[b]{Marco~Bellato}
\author[b]{Antonio~Bergnoli}
\author[a,b]{Matteo~Bolognesi}
\author[a,b]{Riccardo~Brugnera}
\author[c]{Chao~Chen}
\author[d]{Barbara~Clerbaux}
\author[a]{Alberto~Coppi}
\author[b]{Flavio~dal~Corso}
\author[b]{Daniele~Corti}
\author[e]{Jianmeng~Dong}
% ...
\cortext[cor2]{Now at INFN Laboratori Nazionali del Sud, Italy}
\author[a,b]{Chiara~Sirignano}
\author[l]{Monica~Sisti}
\author[b]{Luca~Stanco}
\author[i,j]{Virginia~Strati}
\author[h,g]{Francesco~Tortorici}
\author[h,g]{Cristina~Tuv\'e}
\author[m]{Carlo~Venettacci}
\author[g]{Giuseppe~Verde}
\author[p]{Lucia~Votano}
\address[g]{INFN Sezione di Catania, Catania, Italy}
\address[h]{Universit\`a di Catania, Dipartimento di Fisica e Astronomia, Catania, Italy}
\address[i]{INFN Sezione di Ferrara, Ferrara, Italy}
\address[j]{Universit\`a degli Studi di Ferrara, Dipartimento di Fisica e Scienze della Terra, Italy}
\address[k]{INFN Sezione di Milano e Universit\`a di Milano, Dipartimento di Fisica, Milano, Italy}
\address[l]{INFN Sezione di Milano Bicocca, e Universit\`a di Milano Bicocca, Dipartimento di Fisica, Milano, Italy}
\address[m]{INFN Sezione di Roma Tre e Universit\`a di Roma Tre, Dipartimento di Matematica e Fisica, Roma, Italy}
\address[n]{INFN, Sezione di Milano Bicocca e Politecnico di Milano, Dipartimento di Energetica, Milano, Italy}
\address[o]{INFN Sezione di Perugia e Universit\`a di Perugia, Dipartimento di Chimica, Biologia e Biotecnologie, Perugia, Italy}
\address[p]{Laboratori Nazionali dell'INFN di Frascati, Italy}

\begin{document}

\nocite{*}

\def\sectionautorefname{Section}
\def\subsectionautorefname{Section}

\begin{abstract}
The Jiangmen Underground Neutrino Observatory (JUNO) is a large neutrino detector currently under construction in China. JUNO will be able to study the neutrino mass ordering and to perform leading measurements detecting terrestrial and astrophysical neutrinos in a wide energy range, spanning from 200 keV to several GeV. Given the ambitious physics goals of JUNO, the electronic system has to meet specific tight requirements, and a thorough characterization is required. 
The present paper describes the tests performed on the readout modules to measure their performances. 
\end{abstract}


\maketitle
\section{Introduction} \label{sec:intro}

The Jiangmen Underground Neutrino Observatory (JUNO)~\cite{bib:juno:yb} is a next-generation neutrino experiment under construction in South China, whose aim is to tackle unresolved issues of neutrino physics and astrophysics. The experiment has been proposed with the main goal of determining the neutrino mass ordering (NMO) at 3~$\sigma$ significance in 6 years and providing a measurement of the oscillation parameters with sub-percent precision~\cite{bib:oscillations}. 

The JUNO central detector (CD) is a 20~kton liquid scintillator (LS) in a medium baseline configuration, which is ideal for determining the NMO by studying electron antineutrinos produced by the nearby Yangjiang and Taishan Nuclear Power Plants.
Particle interactions in the LS generate scintillation (dominant) and Cherenkov (sub-dominant, $\leq$~10\%) photons, which are then  
converted into photo-electrons (PEs) by 17612 20-inch Photomultiplier Tubes (PMTs) (\emph{Large-PMTs}) and 25600 3-inch PMTs (\emph{Small-PMTs}).
In addition, 2400 Large-PMTs are installed in the instrumented Water Pool detector in which the CD is immersed.

The initial design of the Large-PMTs electronics~\cite{bib:juno:cd} and the following R\&D program~\cite{bib:elec:bx} have been driven by the main requirement of reconstructing the deposited energy in the LS with an unprecedented energy resolution of 3\% at 1~\si{\mega\electronvolt}~\cite{bib:juno:cd} and a good linearity response (non-linearity $\leq1~\%$) over a wide dynamic range: the average number of PEs produced by a single Large-PMT ranges from 1 PE,
for low energy events, up to thousands of PEs, for showering muons and muon bundles. To ensure accurate vertex and muon track reconstruction, the arrival time of the photo-electrons must be precisely established in both cases.
A basic constraint on the energy resolution arises from the statistics of detected PEs, and it must not be worsened significantly by the effect of the electronics.
Furthermore, the Front End (FE) boards are placed in sealed boxes at a maximum water depth of 43 m, making it impossible to repair or access them after installation. As a result, a high reliability is necessary, with a maximum failure rate of 0.5\% over 6 years of operation.
The latest guidelines for the electronics design can be found in~\cite{bib:el_paper}.

Several tests were performed in order to verify the Large-PMTs electronics specifications, which are required to fulfill the JUNO physics goals.

\section{JUNO Large-PMT electronics}
\label{sec:design}
\begin{figure}[htbp]
\centering
  \includegraphics[width=1\columnwidth]{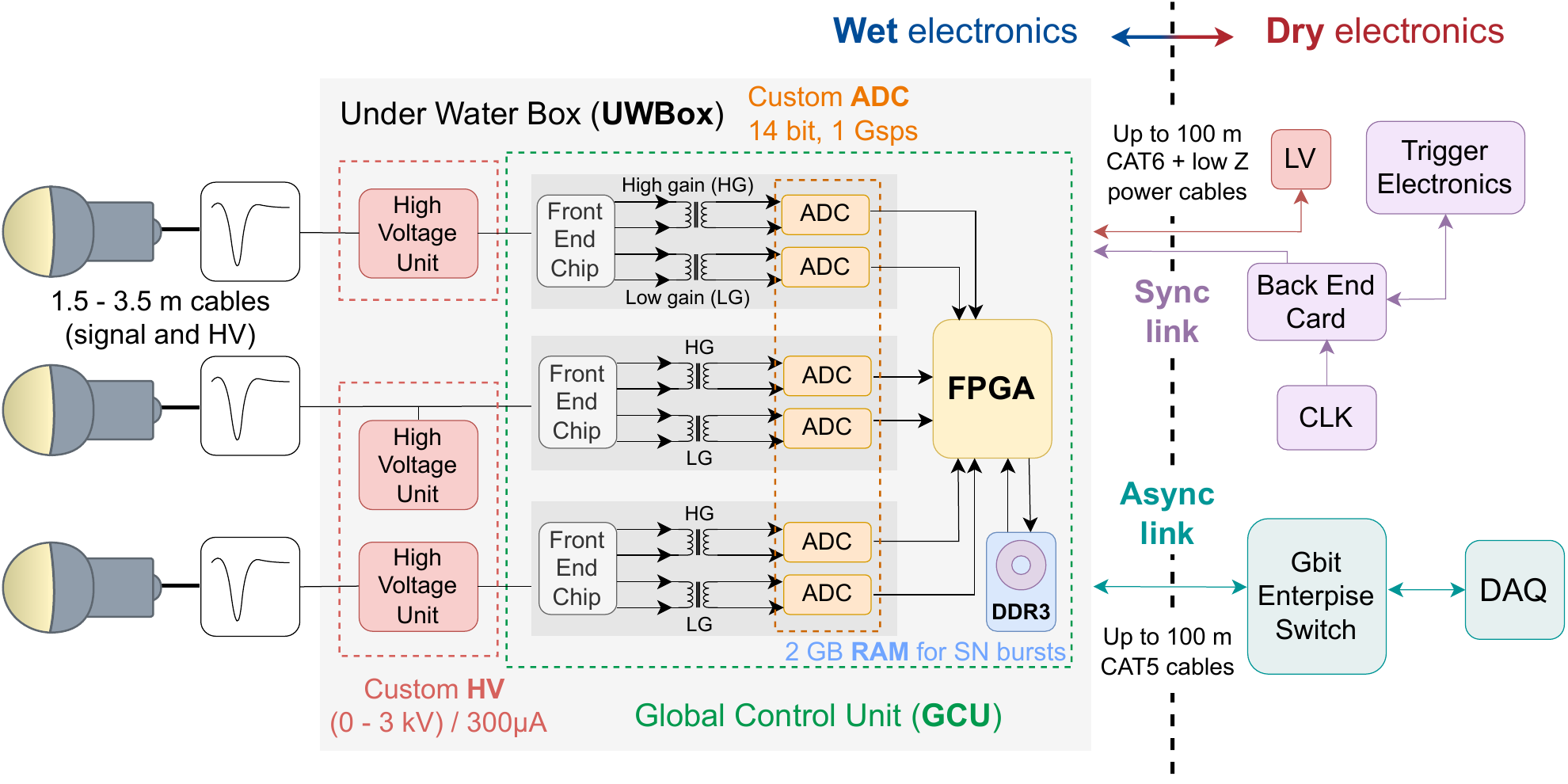}
  \caption{\label{fig:1f3:scheme}JUNO large PMT electronics Read-Out
           electronics scheme. A description of the different parts is
           given in the text.}
\end{figure}

A scheme of the JUNO Large-PMT electronics is given in \autoref{fig:1f3:scheme}~\cite{bib:el_paper}; the design is an optimization of previous developments~\cite{bib:elec:bx}.
The full electronics chain is composed of two parts: the \emph{front-end} (FE), or \emph{wet} electronics, located very close to the PMT output, inside the JUNO Water Pool; and the \emph{dry} electronics, installed in the electronics room of the JUNO underground laboratories, and consisting of the \emph{back-end} (BE), or trigger, electronics and the data acquisition (DAQ) system.
The FE electronics will be installed underwater on the JUNO Steel Truss structure, inside a stainless steel, water-tight box, the so-called Under Water Box (UWbox).
The JUNO detector will be instrumented with 6681 UWboxes, 5878 for the CD and 803 for the Water Pool, as part of the JUNO Veto system.
Three PMT output signals are fed to one UWbox which contains:
\begin{itemize}
\item three High Voltage Units (HVU): programmable modules which provide the bias voltage to the PMT voltage divider. Each HVU independently powers one Large-PMT. The HVUs are mounted on a custom Printed Circuit Board (PCB), the splitter board that provides mechanical stability to the modules, and decouples the PMT signal current from the high voltage.
\item a Global Control Unit (GCU): a motherboard incorporating the
Front-End and Readout electronics components. The three PMT signals
reaching the GCU are processed though independent readout chains.
\end{itemize}
The PMT analog signal reaching the GCU is processed by a custom Front-End Chip (FEC), which duplicates the input signal and amplifies it with a low-gain and high-gain transimpedance amplifiers (see \autoref{fig:1f3:scheme}). The two signals are further converted to a
digital waveform by a 14 bit, 1 GS/s, custom Flash Analog-to-Digital Converter (FADC). 
The usage of two FADCs per readout channel has been driven by the design requirement of providing a wide dynamic range in terms of reconstructed PEs: from \SI{1}{PE} to \SI{100}{PEs} with a \SI{0.1}{PE} resolution (high-gain stream) and from \SI{100}{PEs} to \SI{1000}{PEs} (low-gain stream) with a \SI{1}{PE} resolution~\cite{bib:juno:yb, Liu:2022nhe}.

A Xilinx Kintex-7 FPGA (XC7K160T) is the core of the GCU and allows to further process the digital signal (local trigger generation, charge reconstruction and timestamp tagging) and temporarily store it in a local memory buffer before sending it to the DAQ.
Besides the local memory available in the readout-board FPGA, a 2~GBytes DDR3 memory is available and used to provide a larger memory buffer in the exceptional case of a sudden increase of the input rate, which overruns the current data transfer bandwidth between the FE electronics and the DAQ.
An additional \mbox{Spartan-6}
FPGA (XC6SLX16) is available on the same motherboard.
It implements a 2-port Ethernet hub and a RGMII interface between the
PHY network chip and the Spartan-6 and it also interconnects the Spartan-6
and the Kintex-7. The Spartan-6 FPGA provides an important
failsafe reconfiguration feature of the Kintex-7 by means of a virtual
JTAG connection over the IPbus, removing the need of a dedicated JTAG
connector and cable.

The BE electronics is composed of the following active elements:
\begin{itemize}
\item the Back End Card (BEC) with the Trigger and Time Interface Mezzanine (TTIM)
\item the Reorganize and Multiplex Units (RMU) and the Central Trigger Unit (CTU), which are part of the Trigger Electronics.
\end{itemize}

The PMTs are connected to the UWbox electronics with a \SI{50}{\ohm}, coaxial cable, with a length ranging between \SI{1.5}{m} and \SI{3.5}{m}. The electronics inside the UWbox has two independent connections to the BE electronics: a so-called \emph{synchronous link} (S-link), which provides the clock and synchronization to the boards and handles the trigger primitives, and an \emph{asynchronous link} (A-link) which is fully dedicated to the DAQ and slow-control, or Detector Control System (DCS). These connections are realized using commercially available CAT-5 and CAT-6 Ethernet cables for the A-link and S-link, respectively; the length of the cables ranges between \SI{30}{m} and \SI{100}{m}.
An additional, low-resistance, power cable will be used to bring power
to the electronics inside the UWbox.

The Large-PMT electronics can run with a centralized \emph{global trigger} mode, where the information from the single \emph{fired} PMTs is collected and processed in the Central Trigger Unit (CTU). The latter validates the trigger based on a simple PMT multiplicity condition or a more refined topological distribution of the fired PMTs in JUNO. Upon a trigger request, validated waveforms are sent to the DAQ event builder through the A-link. The IPBus Core protocol is used for data transfer, slow control monitoring, and electronics configurations.

An alternative scheme is possible where all readout boards send their locally triggered waveforms to the DAQ, independently of each other. With this approach, all the digitized waveforms, including those generated by dark noise photo-electrons, will be sent to the DAQ.
\section{Experimental setup overview}
\label{sec:setup}

To validate the full electronics performances, a medium size
setup with 48 independent channels has been built and operated at the
Legnaro National Laboratories (LNL) of the Italian National Institute of Nuclear Physics (INFN).
%
\subsection{Small-scale test setup}\label{sec:lnl_setup}

\begin{figure}[hbtp]
  \centering
  \includegraphics[width=0.95\columnwidth]{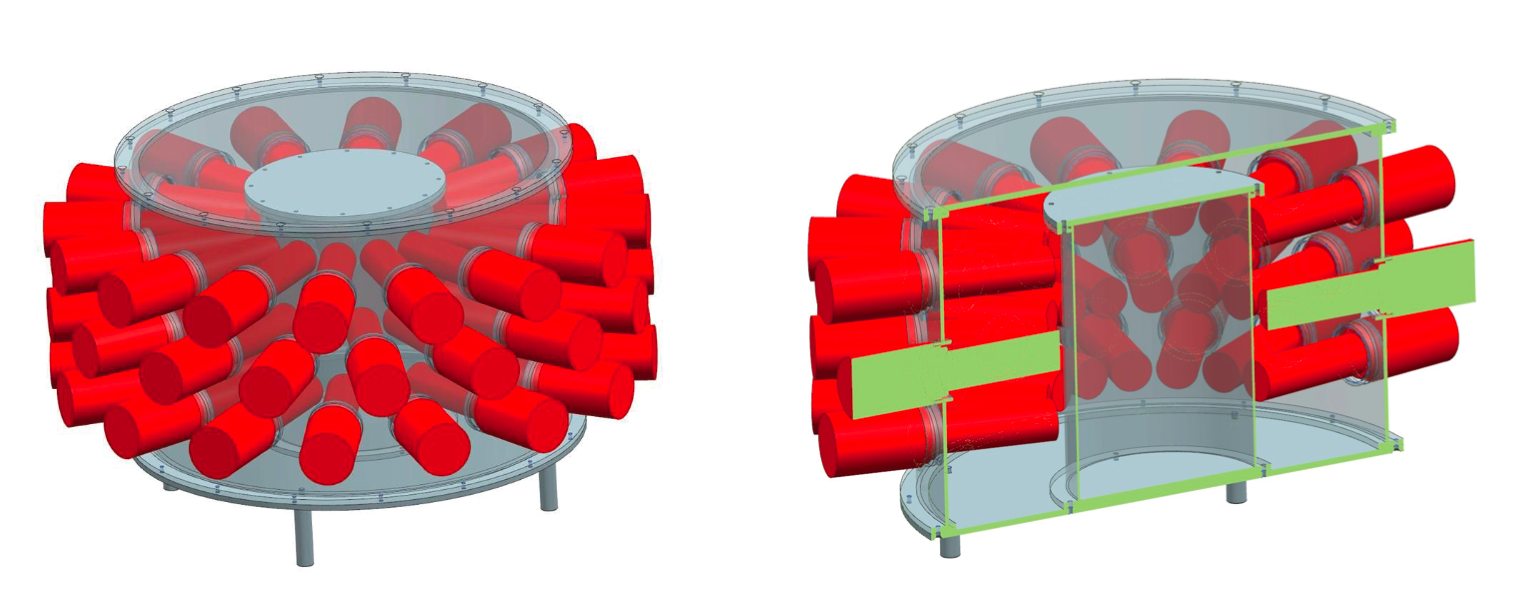} 
  \caption{Mechanical design of the test facility; 
  the internal cylindrical vessel (grey) is surrounded by the 48 PMTs (red); 
  the PMTs are inside the plastic structure (grey), while their bases are outside.}
\label{fig:lnlsetup} 
\end{figure}

\begin{figure}[t]
  \centering
   \includegraphics[width=0.475\columnwidth]{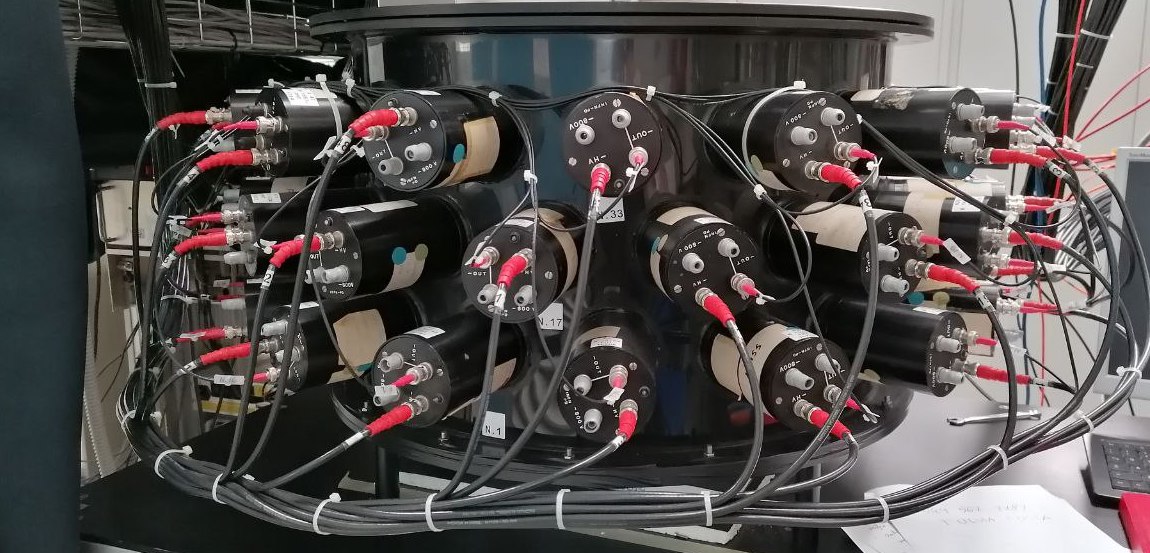}
     \hfill 
     \includegraphics[width=0.47\columnwidth]{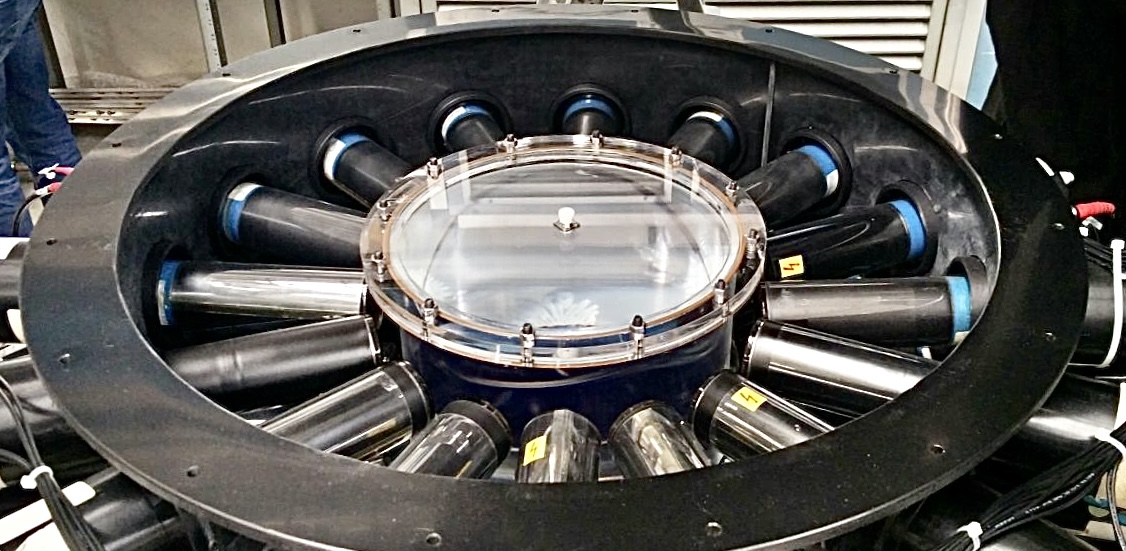}
     \vspace{0.5cm}
    \caption{\label{fig:lavatrice}Picture of the test facility: the black plastic 
    structure and the PMTs base (on the left), the liquid scintillator vessel and the 
    PMTs (on the right) are visible. }
    \label{fig:lnl_testfacility} 
  \end{figure}

The apparatus is composed of a cylindrical acrylic vessel, made of transparent Plexiglass, with inner dimensions of 25~cm diameter
and 35~cm height, filled with about 17~liters of liquid scintillator (LS). The liquid scintillator is composed of a solvent, linear
alkylbenzene (LAB), doped with Poly-Phenylene Oxide (PPO) and p-bis-(o-MethylStyryl)-Benzene (bisMSB), used as wavelength shifter to
match the PMT response; the LS mixture has been optimized using one detector of the Daya Bay experiment~\cite{bib:LS}. 
The LS vessel is inserted in a coaxial larger cylindrical structure
that supports 48 2-inch photomultipliers arranged in three rings, with
16 PMTs each. The inner vessel is surrounded by a black plastic 
structure that supports the PMTs and shields the liquid scintillator 
vessel from the external light. A drawing of the test system 
mechanical design and its realization can be found in \autoref{fig:lnlsetup} and \autoref{fig:lnl_testfacility}, respectively.

The PMTs are Philips XP2020 with custom base, operated with the
photo-cathode at positive high voltage and the anode on
ground\footnote{This is the opposite of what is done in JUNO where the
photocathode is on ground, while the anode is operated at negative high
voltage.}. The PMTs are suitable for high and medium energy physics where 
the number of photons to be detected is very low; moreover, they feature a good linearity, a very low background noise (the typical anode dark noise is up to 900~Hz), and extremely good time characteristics~\cite{bib:XP2020}.

The setup is equipped with ancillary systems (e.g., plastic scintillators to trigger on cosmic muons) that can be exploited to induce signal pulses on the PMTs; specifically, in this paper, we discuss some results obtained using a laser light source introduced inside the liquid scintillator vessel via an optical fiber and a diffuser. It allows to generate narrow pulses and is well suited to investigate the timing characteristics of the electronics.

\subsection{Electronics chain}\label{sec:el_chain}
The electric signal obtained from the collected charge by the PMTs goes through several steps of the electronics chain, before being stored 
on disk. The 48 PMTs are connected, in groups of 3, to 16 GCUs; only 
13 GCUs were available for the measurements described in the following
(i.e., 39 acquisition channels). The electronics chain, the schematic 
description of which is reported in \autoref{fig:elchain}, works according to the following steps:
\begin{figure}[hbtp]
  \centering
  \includegraphics[width = 0.95\columnwidth]{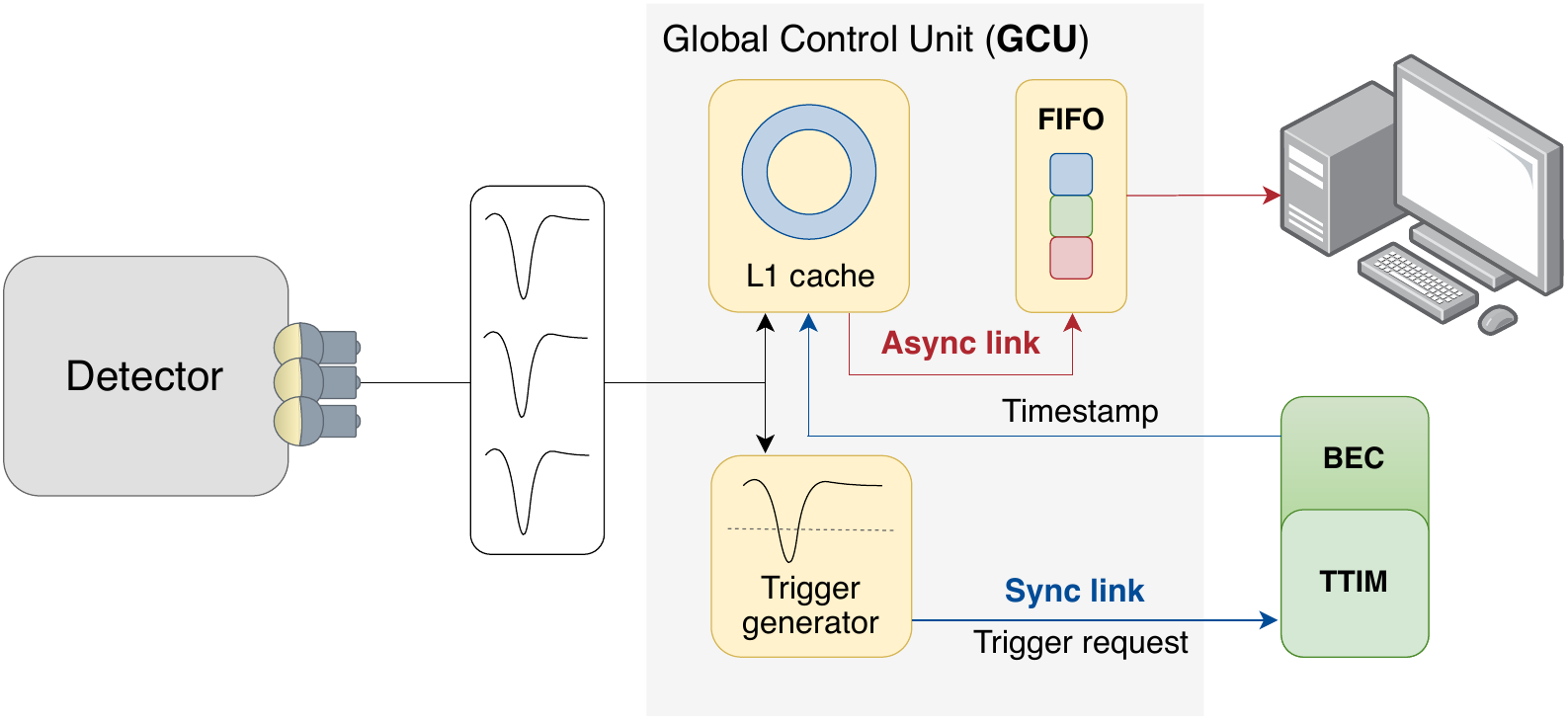} 
  \caption{Electronics chain}
\label{fig:elchain} 
\end{figure}
\begin{itemize}
  \item all three channels acquire their signals concurrently;
  \item inside the GCU, each PMT analog signal is processed by
  a Front-End Chip, which splits the signal into two streams with different gain: a low gain with a dynamic range from \SI{100}{PEs} to \SI{1000}{PEs}, and a high gain with a reduced range from \SI{1}{PE} to \SI{100}{PEs}. The signals are then fed to the FADC;
  \item inside the FPGA the digitized signal is doubled: one of the two signal copies is registered with its GCU timestamp on the L1 cache, while the other is analyzed with a specific threshold trigger algorithm;
  \item  if the signal exceeds a fixed threshold, the GCU sends a trigger request to one BEC which collects the S-link from 48 GCUs. The TTIM then takes a \emph{global trigger} decision, based on the chosen trigger logic, and sends a global trigger validation signal to all connected GCUs; for this step, GCUs and BEC must be properly synchronized in time~\cite{bib:FPGA}. Namely, the global trigger logic can be based either on a logic OR of all 3 channels of a single GCU, or on the multiplicity of the acquired event, i.e., a logic AND between two or more channels, either of the same GCU or of different ones.
  Besides this trigger validation procedure, the system implements an additional external trigger that can be used, for instance, to trigger on different types of events;
  \item after the trigger validation by the BEC, the firmware retrieves the signals with the selected timestamp from the L1 cache and moves them to a First-In-First-Out (FIFO) unit. The content of the FIFO is then transferred to the server through a Gigabit Ethernet switch, where the DAQ program stores the raw data according to a fixed structure. Data transfer is implemented thanks to the IPBus Core protocol~\cite{bib:ipbus}.
  An in-depth investigation of the implementation and performance of the IPBus in the JUNO data acquisition streams can be found in~\cite{bib:ipbus_paper}.

\end{itemize}

The BE presents some differences with respect to the final JUNO back-end chain. Since the detector has been set up during the electronics development phase, different components were still not available or fully functional; therefore, the BE initially included only the BEC and TTIM.
For this reason, a special TTIM FPGA configuration was developed, which included all the basic trigger decision functionalities and the IPBus connectivity. Nevertheless, this temporary dedicated firmware 
includes the back-end module for the IEEE 1588 synchronization protocol~\cite{bib:FPGA}.

\section{Timing synchronization}\label{sec:synchronization}
All of JUNO’s GCUs must be synchronized and aligned within a global time in order to correctly timestamp the triggered events and reconstruct key parameters of the investigated physics process. 
The clock network is based on the \emph{White Rabbit} (WR) standard, 
which exploits the IEEE 1588-2008 Precision Time Protocol (PTP)~\cite{bib:FPGA}; the latter ideally guarantees the synchronization between local (front-end) and global (back-end) clocks inside a window of one clock period. Since the synchronous link protocol rate is 125~Mbps,
the synchronization window is $\pm$~8~ns. In-depth information regarding the aforementioned synchronization process can be found in~\cite{bib:el_paper}.

\begin{figure}[hbtp]
  \includegraphics[width=1\columnwidth]{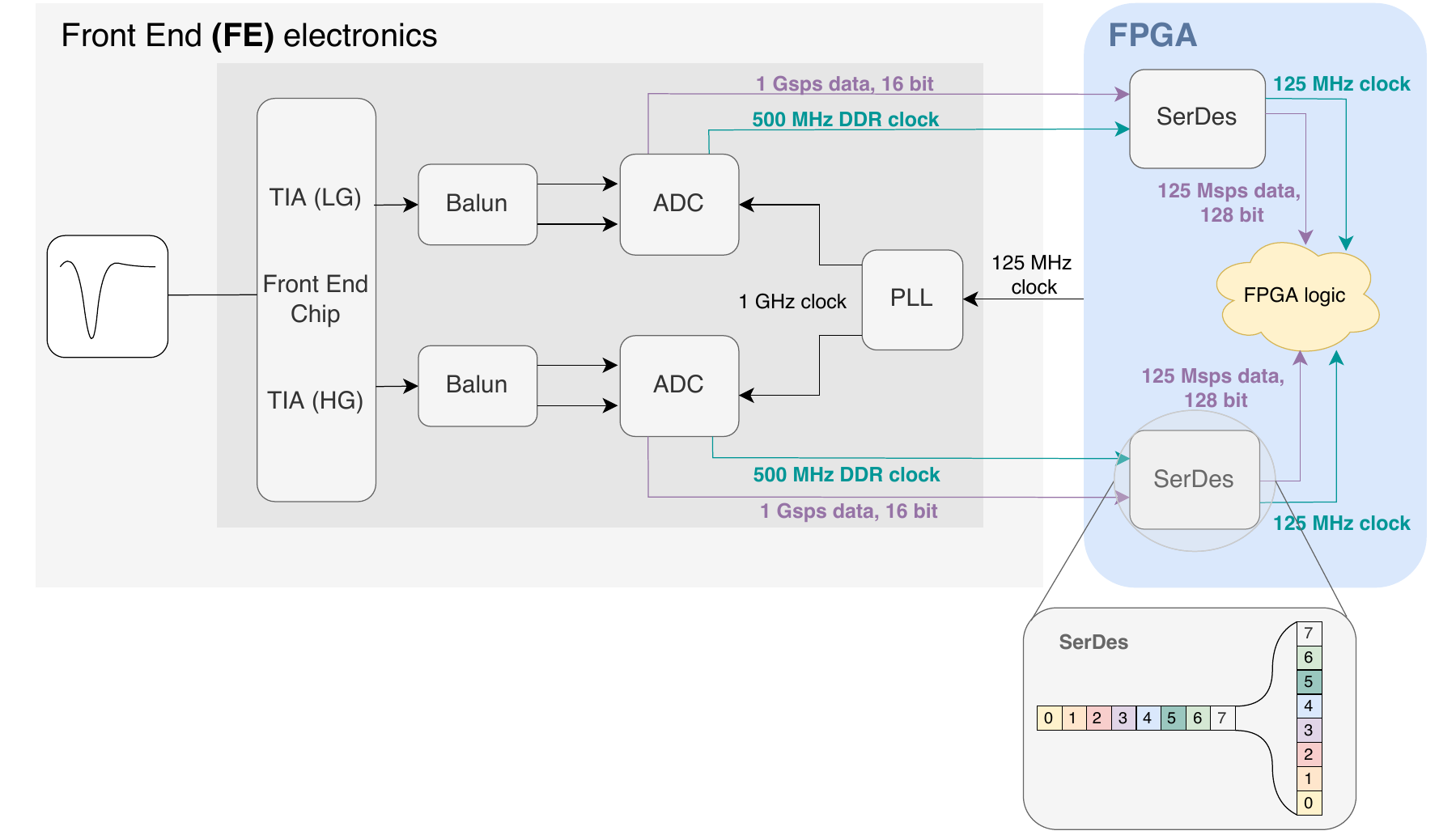}
  \caption{\label{fig:adu}   Block diagram of the FE. The FE electronics provides the FPGA with a 14 bit parallel bus synchronous to its 500~MHz DDR sampling clock. The inset graph represents the SerDes tile behavior in the Input/Output Blocks (IOB): it groups together 8 ADC words, synchronizing them to a system clock of 125 MHz.}  
\end{figure} 

As described in \autoref{sec:el_chain}, the FADC receives the input charge from the PMT, converts it into a voltage, digitizes the waveform, and sends it to the FPGA for further processing. A block diagram of the FE electronics is shown in \autoref{fig:adu}. A stream of 14-bit data, sampled at 1~Gsample/s, is transferred from the ADC to the FPGA, with the data synchronized to a 500~MHz Double Data Rate (DDR) sampling clock. The latter is generated by a Phase-Locked-Loop
(PLL) mounted on the FADC. It receives the system clock
from the GCU and provides a low jitter 1~GHz clock to the ADC. Since the FPGA logic cannot sustain a stream of 1~Gsample/s 14-bit ADC data, a Serializer/Deserializer (SerDes) is used in order to cope with such data rate. The SerDes parallelizes the incoming data 8 to 1, so that 1~Gsample/s 16-bit data (14 bit ADC + 2 bit padding) is parallelized into 125~Msample/s 128-bit data, thereby resulting in an inherent 8~ns phase uncertainty for each channel. The SerDes behavior is depicted in the diagram in
the inset of \autoref{fig:adu}. 

Due to the IEEE 1588-2008 PTP different GCUs have to be synchronized within 16~ns; a second contribution to a potential timing mismatch between two simultaneous PMT pulses detected on two different channels comes from the usage of the SerDes tile to interface with the ADCs. 
Both of these factors introduce a synchronization uncertainty of 16 ns;
while the latter is always present, the first one only affects channels on different GCUs.
Therefore, the timing mismatch is expected to be:
\begin{itemize}
  \item Up to 16~ns for channels of the same GCU. 
  \item Up to 32~ns for channels residing on different GCUs. 
\end{itemize}

\subsection{Synchronization test}\label{sec:sync_tests}

For a proper operation of the system, synchronization among the GCUs has to be stable over time: since it is in principle expected and/or possible to perform the timing realignment at the start of each run, it is sufficient to assure stability within one single run. To evaluate the timing synchronization and mismatch between GCU channels, the 48 PMTs small-scale setup has been used: it is equipped with extremely fast PMTs, suitable for this kind of measurement (they are characterized by 
a time jitter of $\sim$~250~\si{\pico\second} \cite{bib:XP2020}). 
A laser source was employed for the test, the Hamamatsu PLP-10 ultrashort pulsed light source was used: it consists of an M10306 laser diode head and a C10196 controller which provides fast pulses with a FWHM of about 52~ps at a wavelength of 403~nm~\cite{bib:laser}, value that is close to the maximum sensitivity of the PMTs.

First of all, the laser timing was checked by injecting laser pulses into the LS and directly verifying the alignment of the rising edges of signals detected by the single PMTs. The time offsets with reference to a fixed channel are measured by means of an oscilloscope
and range from a maximum of $(2.4 \pm 0.5)$~\si{\nano\second} to a minimum of $(-1.1 \pm 0.2)$~\si{\nano\second}, which can be considered negligible with respect to the expected timing mismatch. Moreover, the PMT hit time is not affected by the laser injection position: indeed no correlation was found with the position of the PMTs inside the three rings.

The test setup for the GCU acquisition is shown in \autoref{fig:elchain}, with the BEC set in external trigger mode, connected to the external trigger output of the laser pulse generator; the laser frequency is set to \SI{2}{\hertz} for the test. Each time the light is emitted, the associated timestamp is received from the BEC via the synchronous link as a global trigger validation, marking the start of the event. 

Due to the aforementioned motivations, waveforms acquired by different channels are not perfectly synchronized but present an offset. With the purpose of evaluating the latter, the time differences between the trigger time\footnote{Timestamp in which the signal reaches an amplitude of 5$~\sigma$ above its baseline.} of the $i$-th channel and the one of a reference channel are evaluated.

\begin{figure}[hbtp] \centering
  \includegraphics[width=1\columnwidth]{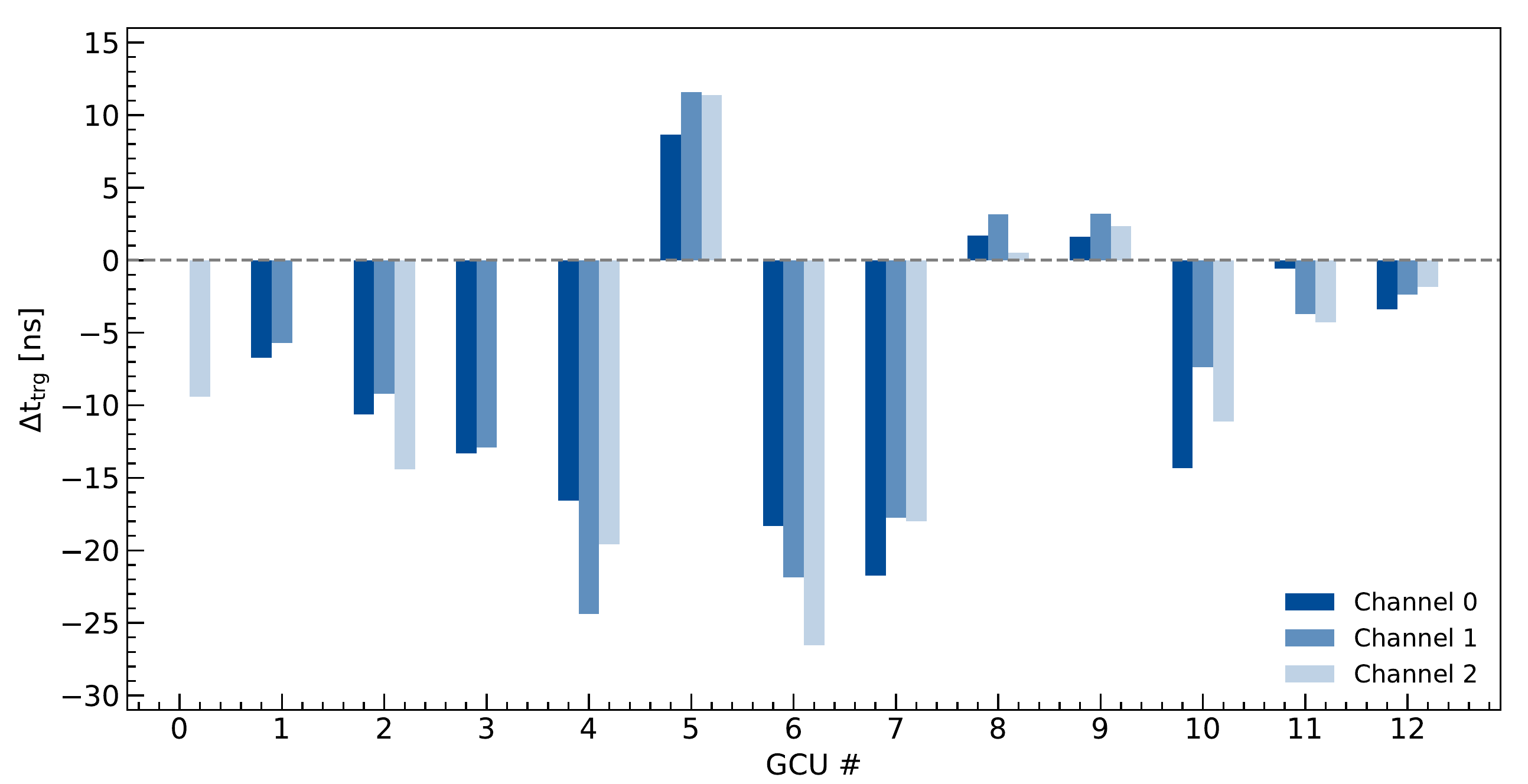}
  \caption{\label{fig:trg_diff} Average time differences $\Delta \mathrm{t}_{\mathrm{trg}}$ for different channels with respect to a fixed one (ch1 of GCU0)}
\end{figure}

\begin{figure}[hbtp] \centering
  \includegraphics[width=1\columnwidth]{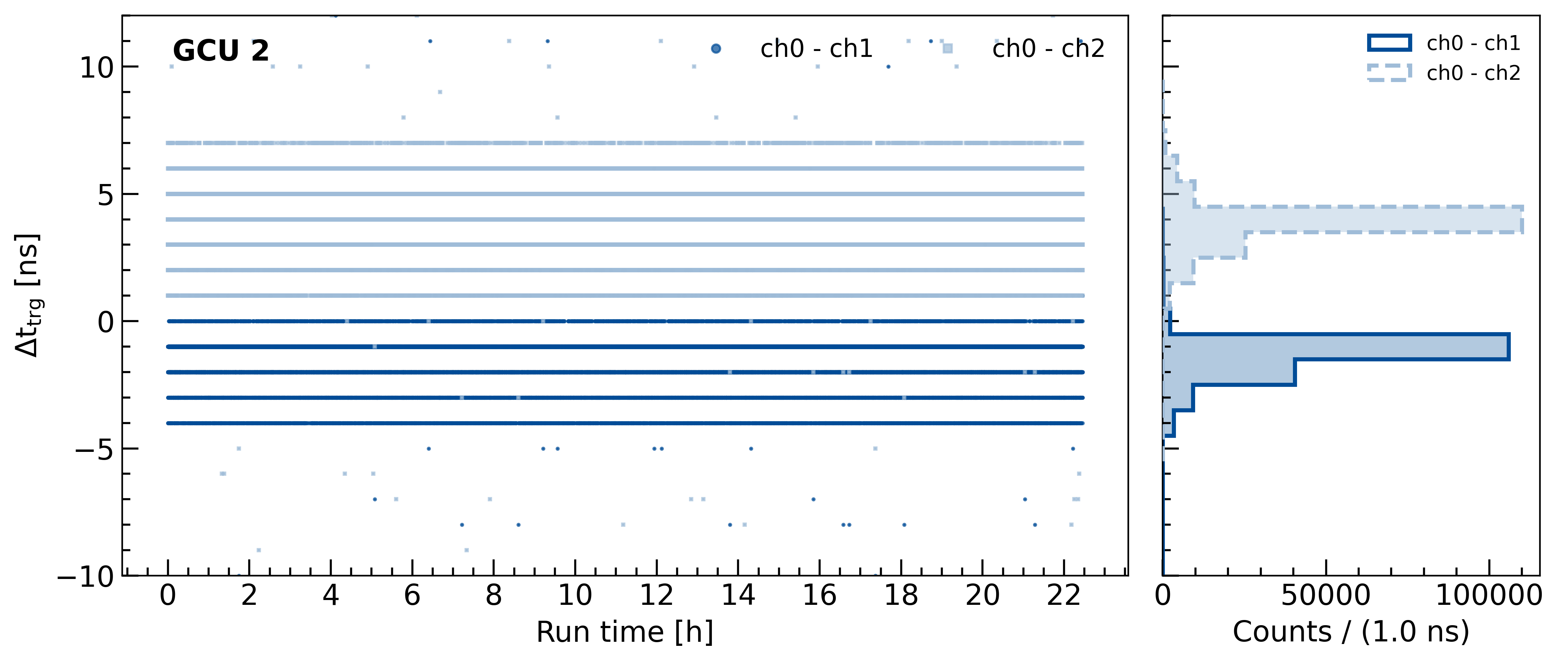}
  \caption{\label{fig:trg_diff_gcu2} Average time differences $\Delta \mathrm{t}_{\mathrm{trg}}$ for two channels of the same GCU, during roughly 22 hours. In this case GCU2 is under study and the differences are calculated for channels 1 and 2 with respect to channel 0. The $\Delta \mathrm{t}_{\mathrm{trg}}$ distributions (on the right panel) are peaked around \SI{-1.5}{\nano\second} and \SI{4}{\nano\second} for channels 1 and 2, respectively. Some outliers, not visible in the plot, are found at more than $3~\sigma$ from the mean value represented only $\sim$~2\% of the sample. The same behaviour is observed for the other channels}.
\end{figure}

The analysis was conducted evaluating the time differences $\Delta \mathrm{t}_{\mathrm{trg}}$ for all 37 channels and choosing channel 1 of GCU0 as reference. All average $\Delta \mathrm{t}_{\mathrm{trg}}$ for the different channels are shown in \autoref{fig:trg_diff}. Firstly, one can infer that channels residing on the same GCU are synchronized within a time interval $\leq$~10~\si{\nano\second}, well inside the expected 16~\si{\nano\second}. This aspect is further illustrated in \autoref{fig:trg_diff_gcu2}, where $\Delta \mathrm{t}_{\mathrm{trg}}$ is evaluated for channels 1 and 2 of GCU2 (used as an example) throughout a one day-long acquisition, and choosing channel 0 as reference.

However, looking at \autoref{fig:trg_diff}, results show that the majority of the signals deviate in time from the reference trigger time by values that range from $\sim$~-26.5~\si{\nano\second} to $\sim$~11.5~\si{\nano\second}, summing up\footnote{In this way it is possible to assess the configuration with the largest possible spread one can find by randomly choosing 2 GCUs for the calculation of $\Delta \mathrm{t}_{\mathrm{trg}}$. }
to $\sim$~38~\si{\nano\second}. The maximum value expected in the global response of the DAQ system is exceeded by roughly 6~ns, which is likely due to the asymmetry of the Cat-5E twisted pairs that provide the physical 
communication link between the FE and the BE. Indeed, the time synchronization performances of the IEEE 1588-2008 PTP protocol 
may be worsened by this aspect~\cite{bib:PTP,bib:PhD-thesis-filippo-marini}. Other deviations may be due to the fact that delays introduced during the PMT photon collection are not taken into account in the $\Delta \mathrm{t}$ evaluation process. However, we preliminarily verified that this effect does not make a significant contribution (up to $\sim$~2.4~ns).

\begin{figure}[hbtp] \centering
  \includegraphics[width=1\columnwidth]{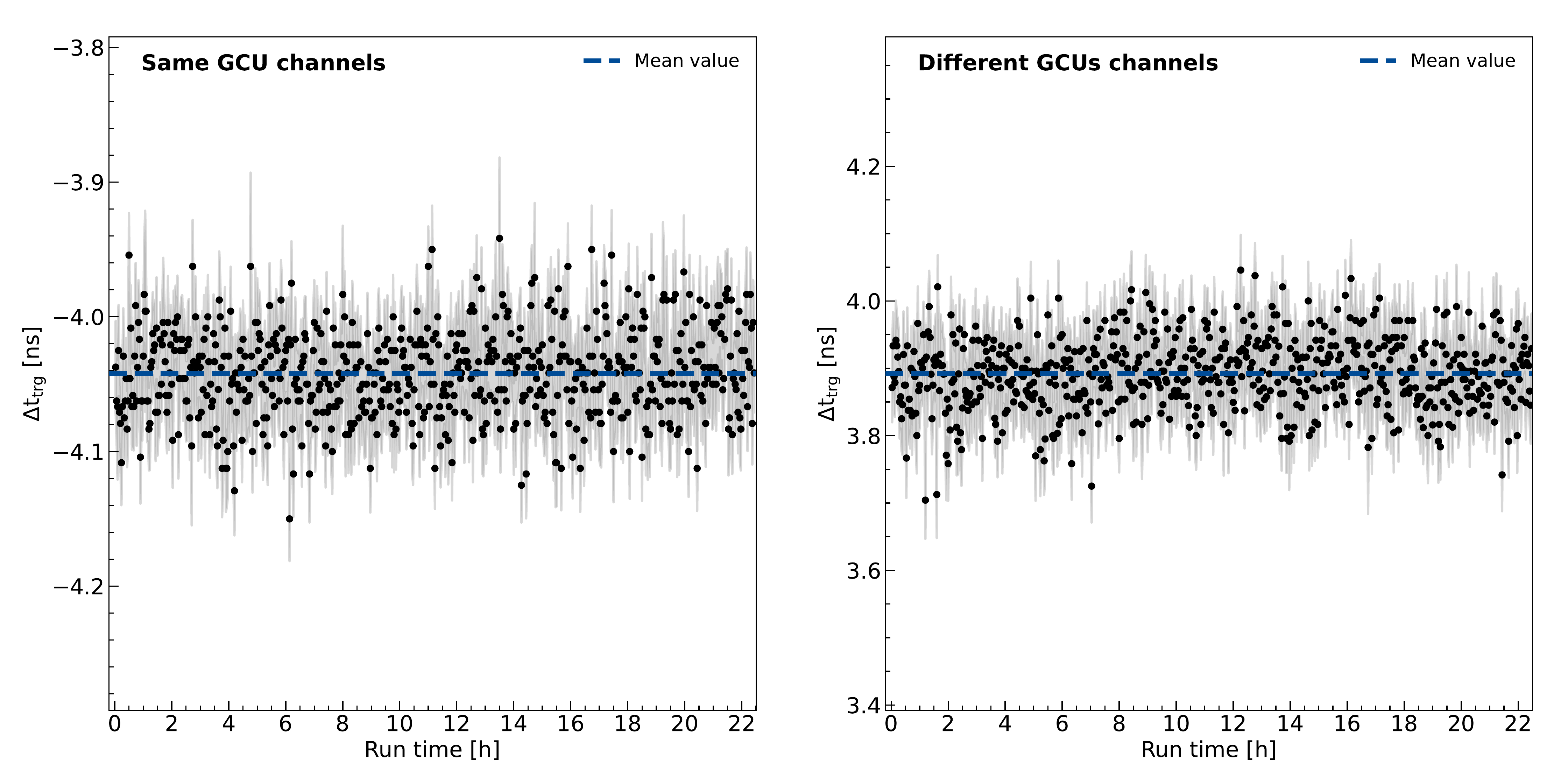}

  \caption{\label{fig:stability_trg} Trigger time difference $\Delta \mathrm{t}_{\mathrm{trg}}$ stability during a one day-long acquisition: left-hand side results for two channels of the same GCU, right-hand side for two channels of different GCUs. Each point represents the mean value of a 2 minutes sample; error bars indicate 
  the standard deviation of the corresponding distribution.}
\end{figure}

Synchronization between different GCUs was further studied by evaluating the trigger time differences over one day-long runs. An example is reported in \autoref{fig:stability_trg}: the channels are synchronized throughout the acquisition period, which means that once the alignment is performed, the difference in the trigger time remains stable within the predicted uncertainty. This experimental outcome shows a good agreement 
with the expectations, thereby playing a key role in the evaluation of the stability in terms of time synchronization.

\section{Charge linearity}\label{sec:linearity}

The determination of the NMO requires the effective energy resolution of JUNO to be better than 3\%~\cite{bib:juno:yb}, an unprecedented requirement in any of the LS-based neutrino experiments. 
Consequently, a comprehensive calibration program~\cite{bib:calibration} is foreseen to account for the intrinsic non-linearity in the scintillation and cherenkov light emitting mechanisms. Nevertheless, the PMT instrumentation and electronics may carry additional \emph{instrumental non-linearity}, namely a non-linear response in the measured charge at a given energy. The latter can vary by more than two orders of magnitude for a single 20-inch PMT. Therefore, a thorough assessment of the linearity response of the electronics is required.

The tests discussed in this section are performed using an external pulse generator, which provides square wave signals with rise and fall times of 2~\si{\nano\second}, a width of 20~\si{\nano\second}, frequency of 1~\si{\kilo\hertz} and amplitudes spanning from roughly 500~\si{\milli\volt} to 3~\si{\volt}. The input charge in pC is computed considering the above-mentioned parameters and an output impedance of 50~\si{\ohm}.

\begin{figure}[t] \centering
  \includegraphics[width=1\columnwidth]{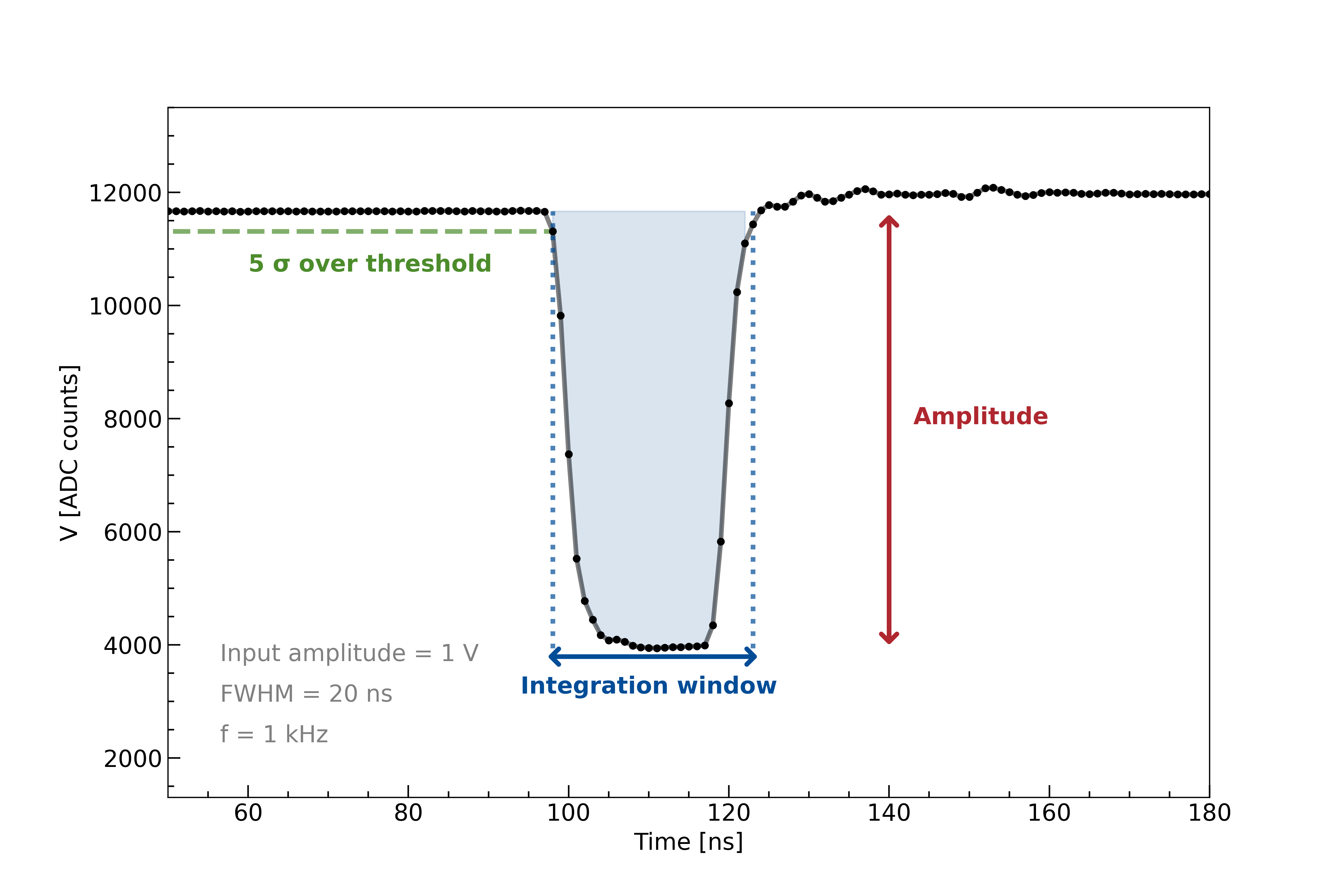}
  \caption{\label{fig:pulser_wf} Example of typical reconstructed waveform obtained using the pulse generator. The integration window is represented by the shaded area. The Full Width Half Maximum (FWHM) and the signal amplitude are indicated as well.   }
\end{figure}

An example of the reconstructed waveform is reported in \autoref{fig:pulser_wf}. 

The output charge ($\mathrm{Q}_{\mathrm{out}}$) is reconstructed by calculating the integral of the signal waveform within 
a fixed time interval, corresponding to a certain number of bins $\mathrm{N}_s$, which can be modified during 
acquisition via the IPbus protocol: 
\begin{equation}\label{eq:q_out}
  \mathrm{Q}_{\mathrm{out}}~[\text{ADC} \times \si{\nano\second}] = \sum^{\mathrm{N}_s}_{i=1} | \mathrm{N}_i - \mathrm{B}| \cdot \Delta \mathrm{t}_i
\end{equation} 
where $\mathrm{N}_i$ is the content of the $i$-th bin in ADC counts, B is the baseline mean value evaluated on a fixed number of samples in the
pre-trigger region (e.g., first 50 samples) and $\Delta \mathrm{t}_i$ is equal to the sampling time (i.e., 1 \si{\nano\second}).
Both baseline and signal time windows are fixed for all events in an acquisition run. In this context, the integration window 
extremes are determined as the time instants when the signal falls below a threshold of 5~$\sigma$ from the baseline. Therefore, for each event, the output charge mean value and its associated uncertainty (standard deviation divided by the square root of the number of pulses) are retrieved.

\begin{figure}[hbtp]
  \includegraphics[width=0.495\columnwidth]{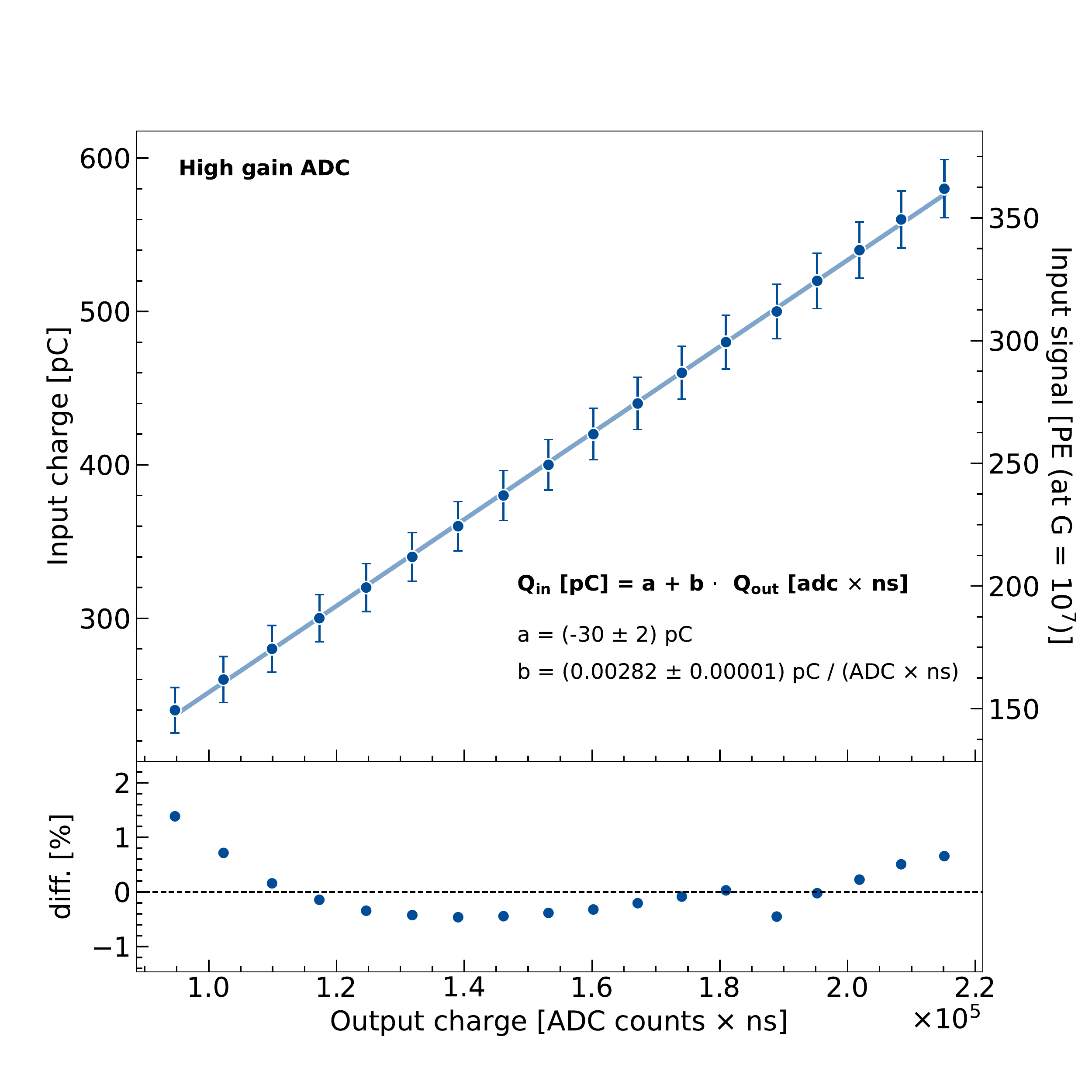}
  \includegraphics[width=0.495\columnwidth]{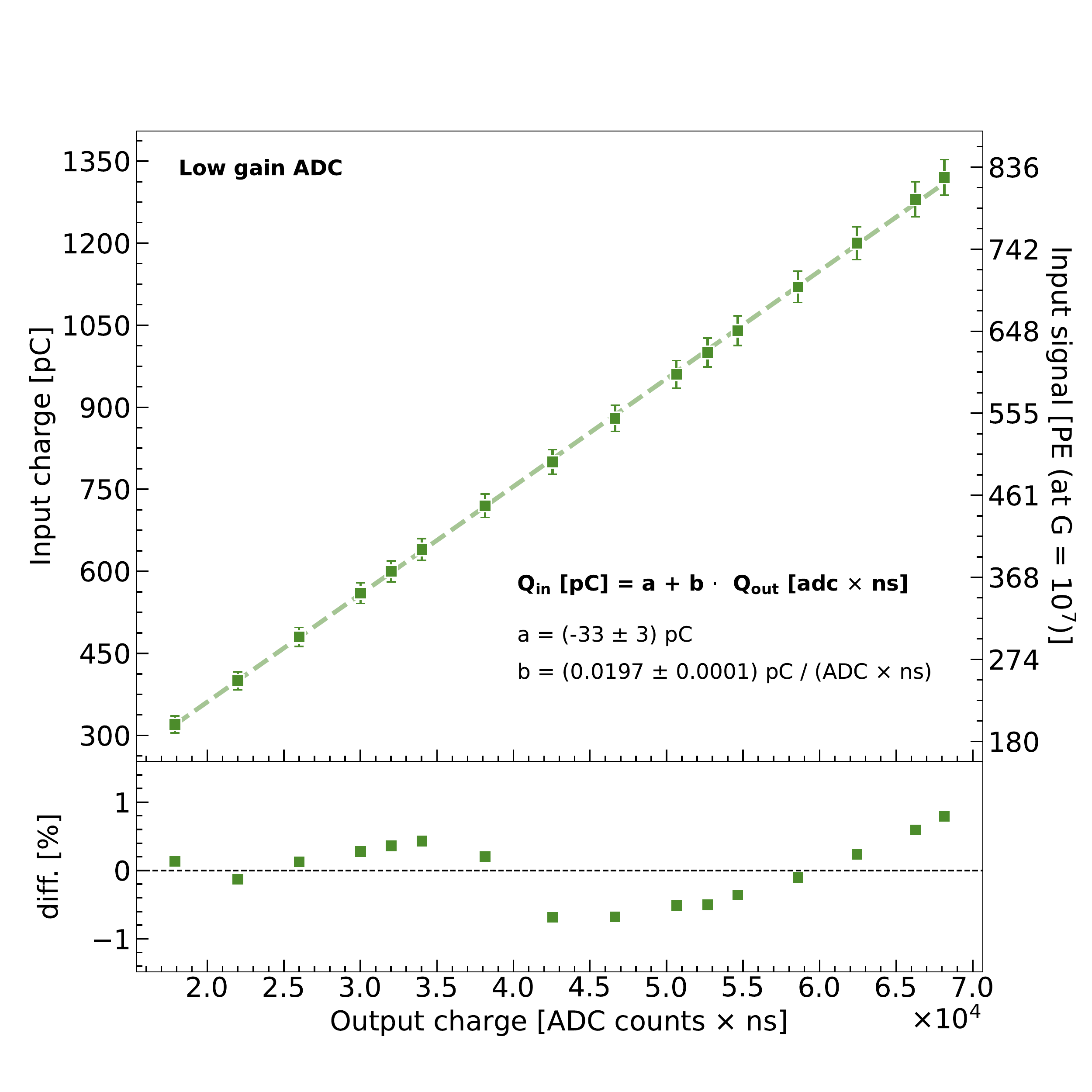}
  \caption{ Linearity of one of the channels for the high gain ADC (on the left), 
  and the low gain ADC (on the right). The top panel shows the calibration curve for the two ADCs with the best fit values;
  in the bottom panel the relative error is reported: diff [\%] = $\frac{y_{data}-y_{th}}{y_{fit}}$}. The uncertainties on the input charge are obtained through 
  propagation, considering the specifications of the employed external pulser.
  \label{fig:linearitycharge}
\end{figure}

In \hyperref[fig:linearitycharge]{Figure 10} the calibration curve for a single channel and both FADCs is reported: the plot shows good linearity and the maximal deviation from a linear fit is $\sim$ 1.4~\% for the high gain ADC and $\sim$ 0.8~\% for the low gain ADC. The systematic trend of the residuals in the bottom panel are most likely due to ADC Differential Non-Linearity.

\begin{figure}[hbtp]
\centering
  \includegraphics[width=1\columnwidth]{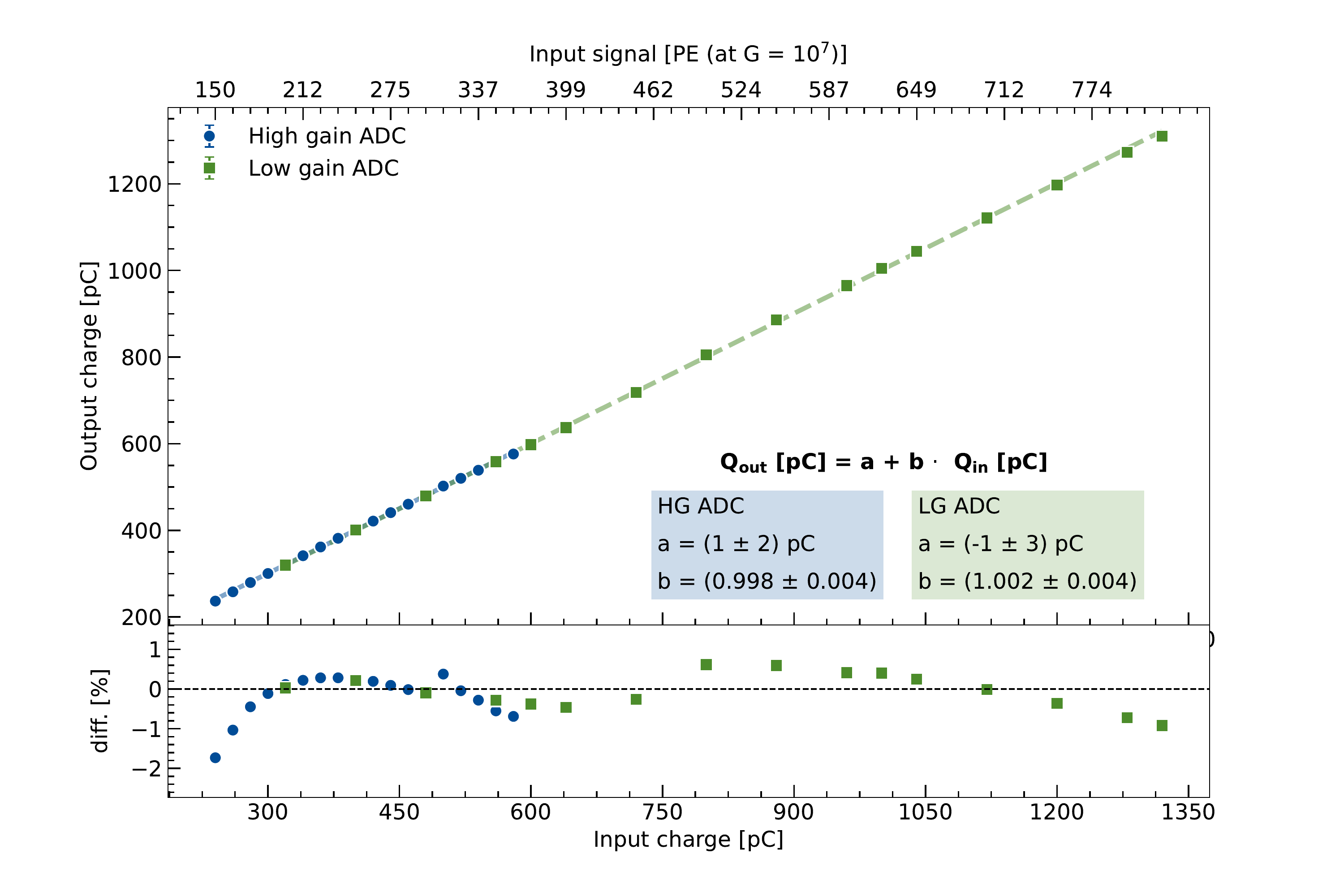}
  \caption{\label{fig:both_adcs_charge} Linearity plot with calibrated data. As reference, the input charge x-scale is also given
   in PE units, assuming a PMT gain of $10^7$.}
\end{figure}

The calibration parameters retrieved from the linear regression are then used to convert the output charge in pC. In \autoref{fig:both_adcs_charge},
the linear response of the electronics is assessed by comparing the input and output charge. As reference, the input charge scale is also given 
in PE units, assuming a PMT gain of $10^7$. In this test, and mainly for the high gain stream ADC, it was not possible to extend the dynamic range to lower PE levels, due to instrumental limitations.\footnote{The poor resolution at low voltage input signal ($\leq$~100-150~\si{\milli\volt}) would have introduced an irreducible instrumental non-linearity.}

The superposition of the high and low gain ADCs curves highlights the goodness of the calibration parameters estimation. The maximal deviation from a linear fit is $\simeq$2~\%. 

\section{UWBox Deep-Water Test}\label{sec:UWboxtest}

A further electronics and mechanical verification has been performed thanks to a collaboration with the \textit{Y-40 The Deep Joy} pool in Montegrotto Terme~\cite{bib:y40}, the deepest thermal water pool in the world, 
with its 42.15 meters in depth. The box stayed underwater at the bottom of the pool for roughly 30 hours. During this time, the FPGA and HVU temperatures were monitored, as well as the baseline average value 
and standard deviation. Since no BEC was used, a modified GCU-standalone version of the firmware that did not 
require the synchronous link was developed. The board was set in auto-trigger mode, where calibration pulses, a feature foreseen in the FEC, were triggered remotely via IPBus.

\begin{figure}[hbtp]
  \centering
  \includegraphics[width=0.495\columnwidth]{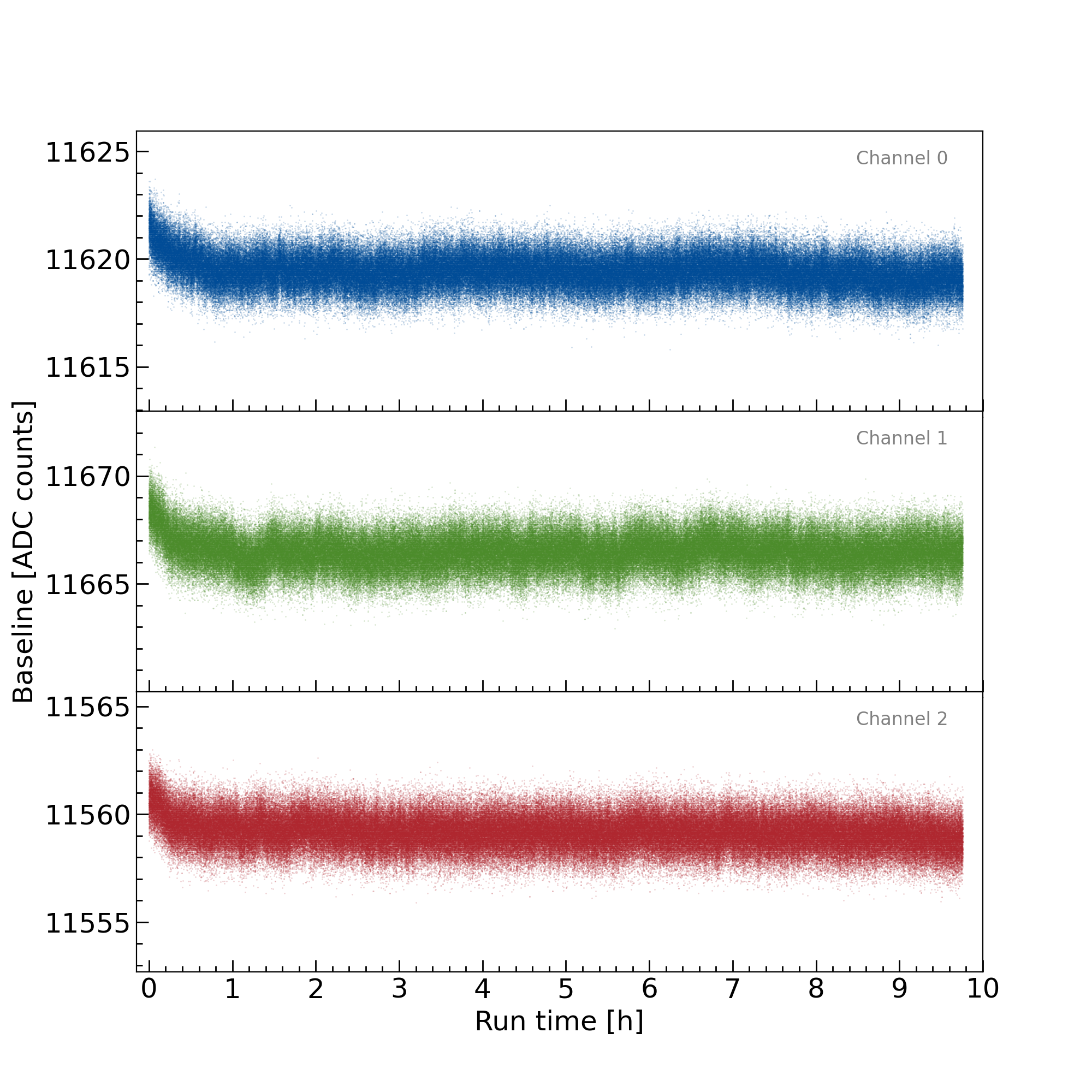}
  \includegraphics[width=0.495\columnwidth]{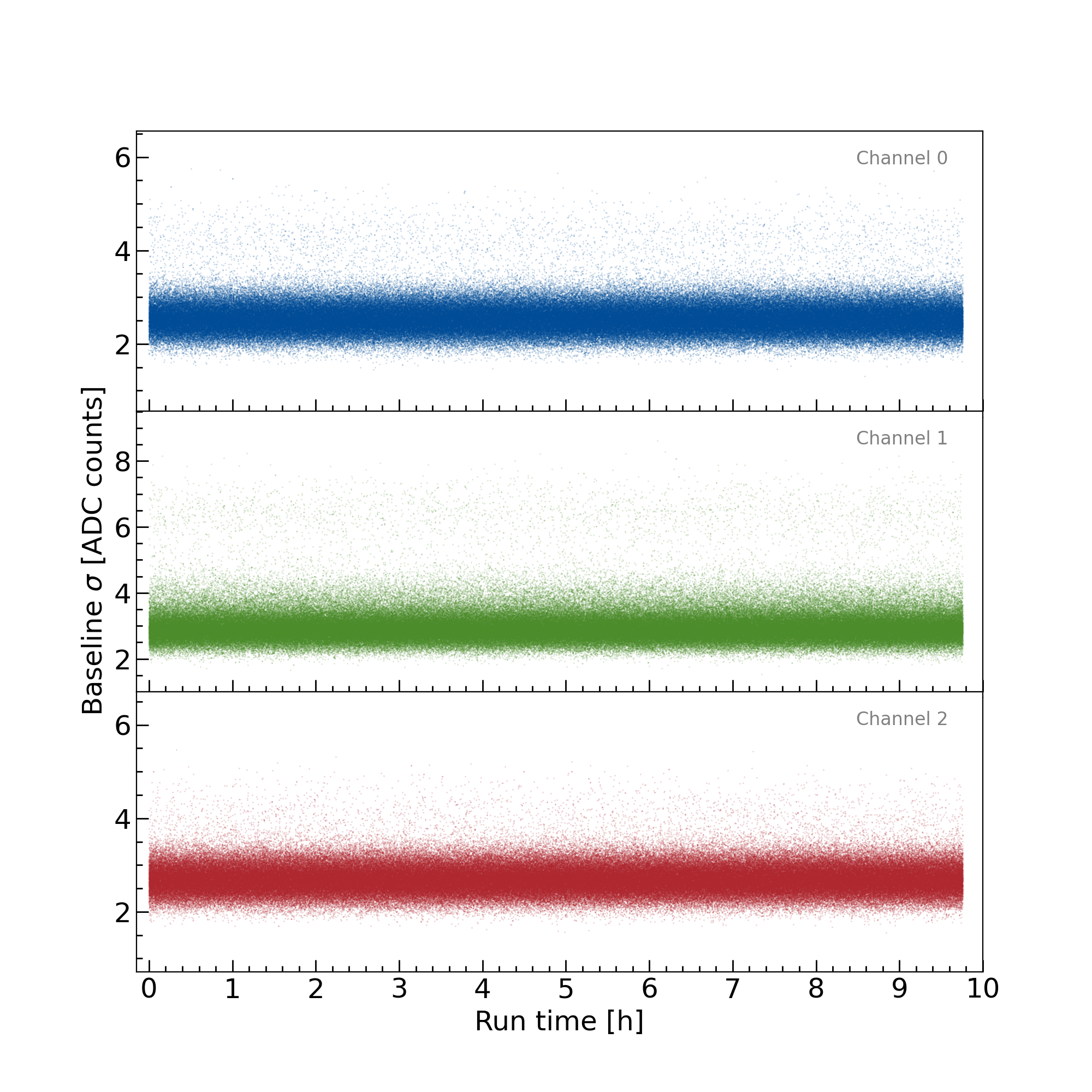}
  \caption{First 10 hours trend of baseline value (left) and baseline standard deviation (right). }\label{fig:bs_y40}
\end{figure}

\begin{figure}[hbtp]
  \centering
  \includegraphics[width=1\columnwidth]{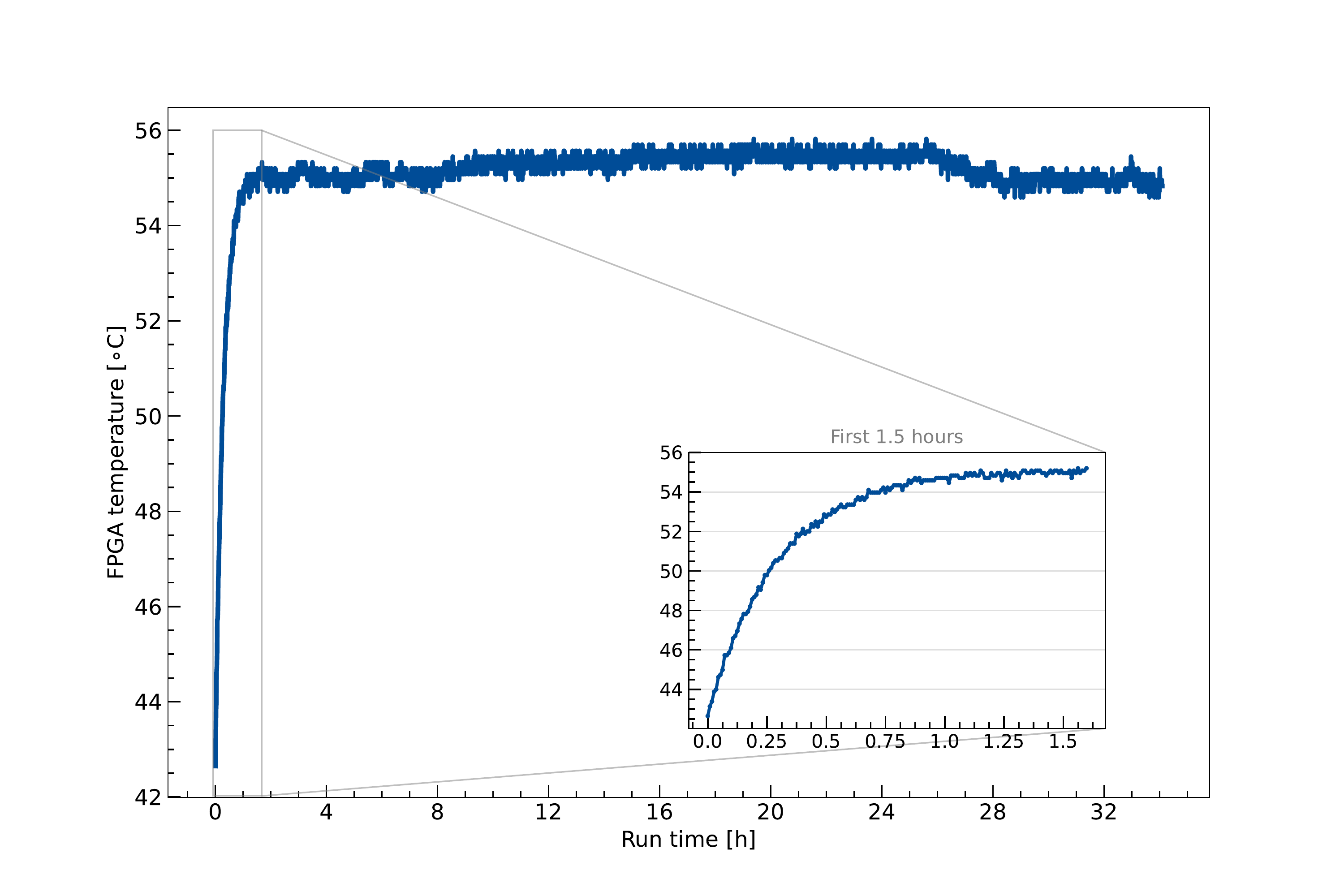} 
  \caption{\label{fig:temp_y40} FPGA temperature monitoring over the $\sim$~32 hours acquisition. The inset plot shows the increasing trend in the first 1.5 hours of the test.The temperature stabilizes after roughly two hours.  }
\end{figure}

In \autoref{fig:bs_y40}, results are reported for all three active channels for the baseline mean (left) and  standard deviation (right). 
Moreover, the FPGA temperature has been recorded: as shown in \autoref{fig:temp_y40}, after a fast initial increase,
it stabilizes at about 55$^\circ$C, with a water temperature of roughly 33$^\circ$C, resulting in a difference of about 22$^\circ$C. Since the water temperature in JUNO is foreseen to be around ($21 \pm 1.4) ^\circ$C~\cite{bib:juno:phys-det}, the FPGA temperature should be below 45$^\circ$C; since the FPGA temperature is more than 15$^\circ$ higher than the one of the board, it falls within the reliability requirements. Indeed, the UWBox cooling should prevent the environment temperature inside the box to raise over 30$^\circ$C~\cite{bib:PhD-thesis-filippo-marini} when in water.

The above-mentioned plots show a correlation between baseline mean value and temperature: as the latter rises, the former decreases, until both of them reach a stable value that remains constant during the acquisition. 
Nevertheless, the water temperature in JUNO is kept constant and therefore no changes in the GCU’s temperature are expected.
The baseline standard deviation value remained unchanged throughout the entire test.

\section{Conclusions}\label{sec:conclusion}
Several tests were performed to assess the performances of the JUNO Large-PMTs electronics.
Synchronization among the GCUs, a key requirement to fulfill the ambitious goals of JUNO, was investigated and monitored over time. The results of the measurements reveal a good agreement with the expected time synchronization performances: the maximum timing mismatch turned out to be 
$\sim$~38~\si{\nano\second}, which exceeds by roughly 6~\si{\nano\second}
the theoretical prediction. However, this discrepancy may be addressed by considering additional sources of timing mismatch, such as the PMT transit time and the cable asymmetry. In terms of trigger stability, an optimal result was obtained: the GCUs proved to remain synchronised during long runs. \\
The linearity response of electronics was also evaluated: the maximal deviation from a linear fit is $\sim$ 1.4~\% for the high gain ADC, in the range (150 - 350) PEs and $\sim$ 0.8~\% for the low gain ADC, in the range (200 - 800) PEs. \\
Additionally, the UWBox was tested $\sim$~40~m underwater, in order to verify its behavior in a JUNO-like environment. The acquired data is consistent with the system's proper operation, and the FPGA recorded temperature complies with the reliability standards.

\section*{Acknowledgements}
Part of this work has been supported by the Italian-Chinese
collaborative research program jointly funded by the Italian Ministry of
Foreign Affairs and International Cooperation (MAECI) and the National
Natural Science Foundation of China (NSFC). The authors are also thankful for the hospitality of the staff of the \textit{Y-40 The Deep Joy} pool.

\bibliographystyle{unsrtnat}
\bibliography{juno_electests}

\end{document}